\documentclass[fleqn,usenatbib]{mnras}

\usepackage[T1]{fontenc}

\DeclareRobustCommand{\VAN}[3]{#2}
\let\VANthebibliography\thebibliography
\def\thebibliography{\DeclareRobustCommand{\VAN}[3]{##3}\VANthebibliography}

\usepackage{graphicx}	
\usepackage{amsmath}	
\usepackage{amssymb}
\usepackage[nopatch]{microtype}
\usepackage{booktabs}
\usepackage{orcidlink}
\usepackage{longtable}
\usepackage{xcolor}
\usepackage{txfonts}
\usepackage{threeparttable}
\usepackage{longtable}
\usepackage{lscape}

\definecolor{DarkBG}{rgb}{0.01 0.1 0.6}

\newcommand{\radmm}{~rad m$^{-2}$~}

\newcommand{\sigrrm}{$\sigma_{RRM}$}

\title[Magnetised Gas in Galaxy Groups and the Cosmic Web with POSSUM]{Probing the Magnetised Gas Distribution in Galaxy Groups and the Cosmic Web with POSSUM Faraday Rotation Measures}

\author[C. Anderson et al.]{
Craig S. Anderson$^{1}$\thanks{E-mail: craig.anderson@anu.edu.au},
N.~M.~McClure-Griffiths$^{1}$,
L. Rudnick$^{2}$,
B. M. Gaensler$^{3,6,14}$,
S.~P.~O'Sullivan$^{4}$,
\newauthor 
S. Bradbury$^{1}$,
T. Akahori$^{5}$,
L. Baidoo$^{6}$,
M. Bruggen$^{7}$,
E. Carretti$^{8}$,
S. Duchesne$^{9}$,
G. Heald$^{9}$,
S. L. Jung$^{10}$,
\newauthor 
J. Kaczmarek$^{11,9}$,
D. Leahy$^{12}$,
F. Loi$^{8}$,
Y. K. Ma$^{1}$,
E. Osinga$^{6}$,
A. Seta$^{1}$,
C. Stuardi,$^{8}$
A. J. M. Thomson$^{9}$,
\newauthor 
C. Van Eck$^{1}$,
T. Vernstrom$^{13}$,
J. West$^{11}$
\\
$^{1}$Research School of Astronomy \& Astrophysics, Australian National University, Canberra, ACT~2611, Australia\\
$^{2}$University of Minnesota, Minneapolis, MN 55455, USA\\
$^{3}$Department of Astronomy and Astrophysics, University of California Santa Cruz, 1156 High Street, Santa Cruz, CA 95060, USA\\
$^{4}$Departamento de Física de la Tierra y Astrofísica \& IPARCOS-UCM, Universidad Complutense de Madrid, 28040 Madrid, Spain\\
$^{5}$National Astronomical Observatory of Japan, 2-21-1 Osawa, Mitaka, Tokyo 181-8588, Japan\\
$^{6}$Dunlap Institute for Astronomy and Astrophysics, University of Toronto, 50 St. George St, Toronto, ON M5S 3H4, Canada\\
$^{7}$Universität Hamburg, Gojenbergsweg 112, 21029 Hamburg, Germany\\
$^{8}$Istituto Nazionale di Astrofisica, Via Gobetti 101, 40129 Bologna, Italy\\
$^{9}$ATNF, CSIRO Space \& Astronomy, Bentley, WA, Australia\\
$^{10}$Department of Physics, University of Oxford, Keble Road, Oxford OX1 3RH, UK\\
$^{11}$Dominion Radio Astrophysical Observatory, Herzberg, National Research Council, Penticton, BC V2A 6J9, Canada\\
$^{12}$University of Calgary, Calgary, AB T2N 1N4, Canada\\
$^{13}$University of Western Australia, 35 Stirling Highway, Crawley, WA 6009, Australia\\
$^{14}$David A. Dunlap Department of Astronomy and Astrophysics, University of Toronto, 50 St. George St, Toronto, ON M5S 3H4, Canada
}

\date{Accepted XXX. Received YYY; in original form ZZZ}

\pubyear{2024}

\begin{document}
\label{firstpage}
\pagerange{\pageref{firstpage}--\pageref{lastpage}}
\maketitle

\begin{abstract}
We present initial results from the Polarisation Sky Survey of the Universe's Magnetism (POSSUM), analyzing 22,817 Faraday Rotation Measures (RMs) with median uncertainties of 1.2 \radmm across 1,520 square degrees to study magnetised gas associated with 55 nearby galaxy groups ($z\lesssim0.025$) with halo masses between $10^{12.5}$ and $10^{14.0}$ M$_\odot$. We identify two distinct gas phases: the Intragroup Medium (IGrM) within 0–2 splashback radii and the Warm-Hot Intergalactic Medium (WHIM) extending from 2 to 7 splashback radii. These phases enhance the standard deviation of residual (i.e., Galactic foreground RM-subtracted) RMs by $6.9\pm1.8$ \radmm and $4.2 \pm 1.2$ \radmm, respectively. Estimated magnetic field strengths are several $\muup$G within the IGrM and 0.1–1 $\muup$G in the WHIM. We estimate the plasma $\beta$ in both phases, and show that magnetic pressure might be more dynamically important than in the ICM of more massive clusters or sparse cosmic web filaments. Our findings indicate that 'missing baryons' in the WHIM likely extend beyond the gravitational radii of group-mass halos to Mpc scales, consistent with large-scale, outflow-driven 'magnetised bubbles' seen in cosmological simulations. We demonstrate that RM grids are an effective method for detecting magnetised thermal gas at galaxy group interfaces and within the cosmic web. This approach complements X-ray and Sunyaev-Zel'dovich effect methods, and when combined with Fast Radio Burst Dispersion Measures, data from the full POSSUM survey --- comprising approximately a million RMs --- will allow direct magnetic field measurements to further our understanding of baryon circulation in these environments and the magnetised universe.
\end{abstract}




\section{Introduction}

Most galaxies in the local universe reside in galaxy groups, which are themselves embedded in the sparse gaseous filaments that form the large-scale structure of the cosmos \citep{HB2014, Bond1996, Eke2004, Rost2024}. Galaxy group environments, particularly through their gaseous, multi-phase, magnetised Intragroup Medium (IGrM), play a pivotal role in mediating energy and matter exchanges between galaxies and the cosmic web, between member galaxies, and  within individual galaxies, through processes such as the transmission of shock waves, turbulence, cosmic ray transport, gas cooling and condensation, and gas heating and redistribution via feedback from galactic winds and active galactic nuclei (AGN) \citep{Lovisari2021, Oppenheimer2021, AG2021, Ayromlou2023}.

However, capturing observational insights into the IGrM's distribution and its state of magnetization remains challenging. The IGrM typically consists of a hot phase ($>10^7$ K) that is dominant within galaxy groups, a warm-hot intergalactic medium (WHIM; $\sim10^6$ K) that becomes dominant near the peripheries of groups, and cooler gas pockets ($\sim10^5$ K) associated with individual group galaxies \citep{Oppenheimer2021}. While X-ray emissions can be used to observe the hot IGrM, they are less well suited to this task than to studies of the intracluster medium (ICM). This is due to the sparser and cooler nature of the IGrM gas, the weakness of bremsstrahlung compared to line emissions, and the complexity of the X-ray emissivity curve at temperatures below 1 keV \citep{Lovisari2021}. Meanwhile, UV absorption techniques provide rich information about cool and warm gas phases in groups \citep{Stocke2019, McCabe2021} but are limited by the density of bright background quasars.

Given these limitations, Faraday Rotation Measures (RMs) emerge as a potent alternative for mapping these elusive gas phases. When linearly polarised electromagnetic radiation traverses magnetised plasma, it experiences rotation in its polarization angle (PA) according to

\begin{equation}
{\rm PA} (\lambda^2) = {\rm PA}_0 + \left[ 0.81 \int_{source}^0 n_e (s) B_\parallel (s)\,{\rm d}s \right] \cdot \lambda^2 \equiv {\rm PA}_0 + {\rm RM} \cdot \lambda^2{\rm ,} \label{eqn:RM}
\end{equation}

where $\lambda$ [m] is the observed wavelength, ${\rm PA}_0$ [rad] is the intrinsic PA (determined by the orientation of the magnetic field in the synchrotron-emitting source), $s$ is the displacement along the line-of-sight to the source [pc], $n_e$ [${\rm cm}^{-3}$] is the electron number density, $B_\parallel$ [$\mu{\rm G}$] is the magnetic field strength along the line-of-sight ($s$), RM [${\rm rad\,m}^{-2}$] is the Faraday Rotation Measure, and the integral is taken along the line of sight from the location of the source to the observer at $s=0$.

By measuring the RM towards ensembles of distant unresolved radio galaxies (so-called 'RM grids'; \citealp{Gaensler2004}), it is possible to detect and map magnetised plasma in a wide variety of foreground environments, limited only by the factors encapsulated in Equation \ref{eqn:RM}, and the sensitivity of radio observations, which determines the density of background RMs (e.g. \citealp{Loi2019}). The advent of sophisticated, wide-field survey instruments such as the Australian Square Kilometre Array Pathfinder (ASKAP), and the Square Kilometre Array (SKA) beyond this, has ushered in a revolution in RM grid science \citep{JH2015,Heald2020}. In particular, ASKAP is undertaking the POlarisation Sky Survey of the Universe's Magnetism (POSSUM), delivering an RM grid covering $2\pi$ steradians with 30--50 RMs per square degree and median RM uncertainty of $\sim1$ \radmm over the next few years \citep{Gaensler2010,Vanderwoude2024,Gaensler2024}, and the SPICE-RACS survey \citep{Thomson2023}, delivering an expected $4\pm2$ RMs per square degree over $3\pi$ steradians even sooner.

A pivotal result from the POSSUM early science phase was the detailed RM grid study of the Fornax Cluster \citep{Anderson2021}, which at $M\sim6^{+3}_{-1}\times10^{13}$ M$\odot$ \citep{Drinkwater2001}, lies on the boundary between a large galaxy group and a poor cluster. \citet{Anderson2021} used an RM grid with 25 RMs per square degree, which although inferior to the full POSSUM survey, improved upon the previous state-of-the-art survey data \citep{TSS2009} by a factor of $\sim25$. In doing so, they uncovered a substantial, previously undetected reservoir of magnetised plasma extending well beyond the detectable X-ray emission from the cluster, providing insights into the dynamical interactions and shock systems induced by merging substructures. This work demonstrated that RM grids could effectively probe gas in halos less massive than large galaxy clusters, where their value for probing the ICM is already well-established \citep{Clarke2001}. Motivated by these findings, this paper reports the results of our search for signatures of magnetised gas in even lower mass halos, both within and beyond the gravitational boundaries of a diverse sample of galaxy groups, using initial RM data from the full POSSUM survey currently in progress. It is organised as follows: Section 2 details the POSSUM observations and our methods; Section 3 presents our analysis and findings; Section 4 discusses these in the context of current astrophysical understanding. Given our focus on the local universe ($cz<10,000$ km/s), we introduce cosmological parameters only where necessary.

\section{Methods}

\subsection{POSSUM observations}\label{sec:obs}

Our study uses 38 out of the planned 853 pointings for the main `Band 1'\footnote{POSSUM observes in ASKAP’s band 1 (800–1088 MHz) commensally with EMU \citep{Norris2011}, but is supplemented by Band 2 (1296–1440 MHz) data from WALLABY where available \citep{Koribalski2020}.} POSSUM survey \citep{Gaensler2024}, each observed for ten hours and covering $\sim40$ square degrees, totaling $\sim$1520 square degrees of sky. These observations were made with the ASKAP radio telescope \citep{Hotan2021} from August 2021 to August 2023 during standard survey operations. The observations recorded full polarisation information over 800-1088 MHz, achieving typical sensitivities of 25--30 $\mu$Jy beam$^{-1}$ in Stokes I \citep{Norris2021,Vanderwoude2024} and 17 $\mu$Jy beam$^{-1}$ in each of Stokes Q and U \citep{Gaensler2024}. Further details of the polarimetric calibration and imaging with ASKAP can be found in \citet{Chippendale2019,Thomson2023,Gaensler2024}. Each observation (or scheduling block; SB) underwent standardised validation procedures described in \citet{Gaensler2024} for polarisation-specific data products and additionally, those described in \citet{Norris2021} for total intensity data products. We excluded SBs that failed these tests. Table 1 summarises our observations.

\begin{table*}
\caption{Summary of observations}
\begin{threeparttable}
\centering
\begin{tabular}{ll}
\hline
\hline
No. of ASKAP pointings & 38 \\
SBIDs included in this work & 31375, 33370, 33460, 33482, 33509, 33553, 43137, 43237, 43773,\\ 
 & 45761, 45781, 45811, 46925, 46943, 46946, 46951, 46962, 46966, 46971,\\ 
 & 46976, 46982, 46984, 46986, 47034, 47130, 47136, 49990, 49992, 50011,\\ 
 & 50230, 50413, 50415, 50423, 50538, 50787, 51431, 51574, 51818
\\
Total period of observations & 28-08-2021 to 03-08-2023 \\
Typical full-band sensitivity in Stokes $Q$, $U$ $^\dagger$ & 17 $\muup$Jy beam$^{-1}$ / Stokes \\
Total sky coverage & $\sim1520$ deg$^2$ \\
Typical Shortest baseline & 22.4 m \\
Typical Longest baseline & 6.4 km \\
Typical Angular resolution (robust $=-0.5$) & 13$\times$17 arcsec \\
Typical Largest recoverable angular scale$^{\ddagger\ddagger}$ & 30 arcmin\\
Frequency range (central frequency) & 800--1088 MHz (288 MHz)\\
$\lambda^2$ range & 0.076--0.141 m$^2$ \\
RM Spread Function FWHM $\textsuperscript{a}$ & 59 rad m$^{-2}$ \\
Largest recoverable RM-scale$^{\ddagger\ddagger}$ $\textsuperscript{a}$ & 37 rad m$^{-2}$ \\
Largest recoverable |RM| $\textsuperscript{a}$ & 8100 rad m$^{-2}$ \\
No. of RMs in final sample & 22,817 \\
Median RM uncertainty & 1.2 rad m$^{-2}$ \\
\hline
\end{tabular}
\begin{tablenotes}
\small
\item $^\dagger$ Measured per Stokes parameter in multi-frequency synthesis images generated with a Briggs' robust weighting value of -0.5. $^\ddagger$ At the centre frequency of the band. $^{\ddagger\ddagger}$ At greater than 50\% sensitivity. $\textsuperscript{a}$ Calculated using RM-Tools version 1.3.1\footnote{https://github.com/CIRADA-Tools/RM-Tools}\citep{Purcell2020}, quoted to 2 significant figures.
\end{tablenotes}
\end{threeparttable}
\label{tab:obsdeets}
\end{table*}

\subsection{The RM sample}\label{sec:prefilteringRMs}

We extracted RMs by further processing the basic \textsc{ASKAPsoft} pipeline~\citep{GW2019} outputs. We note that this does not represent the final POSSUM survey workflow and its associated pipeline data products, which were not available for a significant portion of the sky at the time this publication was prepared. The \textsc{ASKAPsoft} pipeline uses RM synthesis \citep{BdB2005} to calculate Faraday Dispersion Functions (FDFs) from Stokes $Q/I$ and $U/I$ spectra (formed by channel-wise division) extracted from source components identified in total intensity mosaics of each field using {\sc Selavy} \citep{Whiting2012}. We downloaded raw FDFs from the CSIRO ASKAP Science Data Archive\footnote{https://data.csiro.au/domain/casda} (CASDA; \citealp{Huynh2020}) for the SBs in Table 1, then used {\sc RM-Tools}\footnote{RM-Tools version 1.3.1; https://github.com/CIRADA-Tools/RM-Tools} \citep{Purcell2020} to extract RMs, polarised signal-to-noise ratios, and their uncertainties for all radio source components, which are calculated using a quadratic fit to the main peak of the FDF.\\

\subsection{The galaxy group sample}\label{sec:groupcats}

We cross-matched our RMs against the combined catalogues of \citet{MK2011} and \citet{Tully2015}, hereafter referred to as the `MKT' catalogue of galaxy groups. This comprehensive catalogue classifies galaxies based on parameters including mass, recession velocity (`cz'), and proximity to neighbouring galaxies. The \citet{MK2011} catalogue covers the radial velocity range of $0<{\rm cz}<3500$ km/s, while the \citet{Tully2015} catalogue spans the range $3000<{\rm cz}<10000$ km/s. Both catalogues provide integrated K-band luminosity estimates for member galaxies, enabling us to estimate the mass and characteristic radius of each galaxy group. The grouping procedure relies solely on these observables and covers nearly the entire celestial sphere, so is well-suited for our study. The MKT catalogue encompasses a wide range of group mass, from approximately $1\times10^{11}$ M$_\odot$ (an order of magnitude less massive than the Milky Way) to roughly $1\times10^{15}$ M$_\odot$ (the Coma galaxy cluster).

\subsection{RM-Group impact parameters and associations}\label{sec:groupoffsets}

Our goal is to associate each POSSUM RM with its nearest MKT galaxy group by calculating the RM impact parameter for each group (and to exclude RMs from subsequent analysis that are most closely associated with clusters, or fail to meet quality control criteria, as discussed in Section \ref{sec:finalsample}). 

Since the MKT groups have a wide range of distances and masses, and the radius of the IGrM will depend on these properties, we need to establish a fiducial radius to scale the RM impact parameters. To do so, we use the integrated K-band luminosities listed in the MKT catalogue to calculate the group halo mass ($M_\text{group}$) using Equation 7 in \cite{Tully2015}:

\begin{equation}
M_\text{group}=32.25\times10^{10}(L_{10})^{1.15}~\text{M}_\odot,
\end{equation}

\noindent where $L_{10}$ denotes the $K$-band luminosity in units of $10^{10}~L_\odot$, assuming $H_0=75$ km/s/Mpc. Subsequently, we estimate the splashback radius ($R_{spr}$) in Mpc for each group, which represents the region enclosed by the apoapses of gravitationally accreted galaxies following their initial transit through the group halo. The splashback radius is known to correspond to a physical, observable cusp, beyond which matter density decreases rapidly with the projected radius (e.g., \citealp{DK2014,Adhikari2014,More2015}). This relationship can be expressed as:

\begin{equation}
R_{spr}=0.216\times\left(\frac{M_\text{group}}{10^{12} ~\text{M}_\odot}\right)^{1/3}~\text{Mpc}.
\end{equation}

Given that the MKT catalog provides group distances, we can convert $R_{\text{spr}}$ to an angular scale on the sky denoted by $\theta_{\text{spr}}$. We can also calculate the angular impact parameter  denoted $\beta_{\text{RM}}$ from the center of the group for each RM, assuming the center to be the position of the most mass-dominant member galaxy. Using the splashback radius as a fiducial point to scale the RM-group offsets, we then calculate the scaled impact parameter $\xi = \beta_{\text{RM}} / \theta_{\text{spr}}$. $\xi$ thus represents the distance of the RM measurement from the nearest galaxy group in multiples of the splashback radius of that group.

\subsection{Galactic RM foreground corrections}\label{sec:RM foreground}

To isolate extragalactic contributions in our POSSUM RM measurements, we must subtract the Galactic RM foreground to derive residual RMs (RRMs) reflecting extragalactic Faraday rotation  (\citealp{Kim1991}). There are many possible approaches to estimate the foreground. Models such as those by \citet{Hutschenreuter2022} provide one option, but are based on low-resolution RM data (about 1 RM per square degree) and so are not expected to adequately capture the fine-scale Galactic variations that our high-resolution POSSUM data (at least 30 RMs per square degree) can constrain.

On the other hand, L. Jung (personal communication, April 27, 2024) noted that galaxy group halos might contribute spatially-correlated RMs of several \radmm up to distances of $r_{200}$, as shown in IllustrisTNG50 data \citep{Nelson2019}. Observational support for this comes from \citet{Anderson2021}, who documented coherent IGrM/ICM-based RM structures extending out to the virial radius in the Fornax cluster---effectively a large galaxy group---over scales exceeding one degree. This implies that galaxy group halos can significantly influence the local RM field on scales that might otherwise be assumed to be Galactic foreground structure. Clearly, it is critical to avoid subtracting the very signal we wish to study. \citet{Anderson2021} addressed this by fitting a linear 2D RM foreground model over a single ASKAP field, balancing the influence of the smaller cluster RM structures with numerous `off-source' RMs from the surrounding area. However, this method is unsuitable for the present stacking experiment, not least because it could create discontinuities in RRMs across ASKAP tile boundaries.

These challenges underscore the ongoing complexity of Galactic RM foreground removal, which is a topic of active research for next-generation high-density RM grid experiments like POSSUM (\citealp{Khadir2024inprep}). For our purposes, we employ an aperture-based method to subtract the foreground while retaining group-based Faraday rotation signals:

\begin{enumerate}
    \item Exclusion Zone: We set a 0.4-degree exclusion radius around each RM source to prevent subtraction of coherent IGrM-based signals.
    \item Median Estimation: We calculate the median RM of the next-nearest 40 RMs surrounding the source, using this as the local Galactic foreground estimate.
    \item Foreground Subtraction: We subtract this estimated foreground from the source’s RM to obtain the RRM.
\end{enumerate}

The outer bound of the annulus encompassing these 40 RMs typically extends between 0.7 degrees (10th percentile) and 1 degree (90th percentile), with an average radius of 0.8 degrees (50th percentile). We chose a 0.4-degree exclusion radius because it (a) corresponds to the width of coherent RM features observed in the Fornax cluster \citep{Anderson2021}, (b) aligns with the splashback radius of the smallest groups in our sample, and so treats all groups consistently on this basis, and (c) corresponds to the approximate scale where Galactic foreground contributions appear to drop below intrinsic source contributions at mid-Galactic latitudes (\citealp{Anderson2015, Vanderwoude2024}).

To quantify the uncertainty in the foreground subtraction, we used a bootstrap resampling technique where for each RRM value we derive, we:

\begin{enumerate}
    \item resample the RMs in the surrounding annular region 1000 times, and calculate the median RM for each sample.
    \item compute the standard deviation of the bootstrap medians
    \item add this bootstrap-derived uncertainty with the observed $1\sigma$ RM uncertainty in quadrature to obtain the total uncertainty for each corrected RRM.
\end{enumerate}

\noindent The median $1 \sigma$ uncertainty of the Galactic foreground RM derived in this way is 1.2 \radmm, and the median $1 \sigma$ total RRM uncertainty is 2.0 \radmm.

\begin{figure}
\centering
\includegraphics[width=0.48\textwidth]{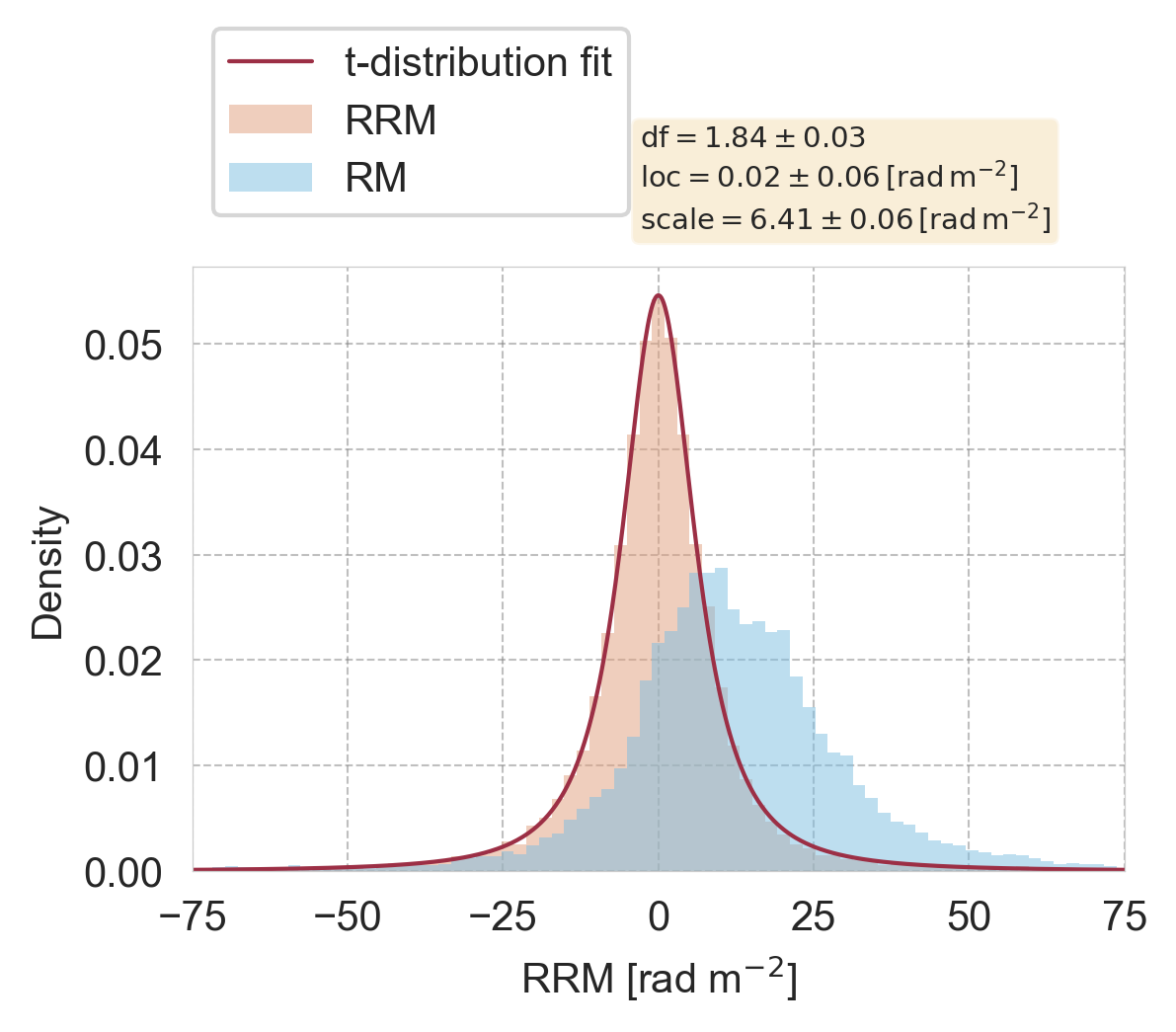}
\caption{Histograms of RM (blue) and RRM (orange) for our final sample. The bin values for RRMs have been fitted with a Student's t-distribution (red line), as discussed in the main text. The best-fit parameter values are listed in the yellow text box.}
\label{fig:RRM_histogram_with_fit}
\end{figure}

Figure \ref{fig:RRM_histogram_with_fit} illustrates a histogram of uncorrected RMs versus RRMs, with RRMs well-fit by a Student t-distribution with df=1.84±0.03, location=0.02±0.06 \radmm, and scale=6.41±0.06 \radmm. The df value indicates tail-heavy distribution with outliers, as expected (\citealp{TSS2009,Schnitzeler2010,AG2021,Thomson2023}). The scale value closely aligns with existing measurements of RM dispersion in extragalactic point sources \citep{Taylor2024, Schnitzeler2010}. Variations in the exclusion radius (0.2 to 1 degree) and the number of RMs (20 to 50) showed only minor effects on these results and those presented in Section \ref{sec:RRMvsRadius}, in the latter case mainly affecting the DC offset of the RM dispersion curves we calculate (by up to 1 \radmm) but not the `stepped' RM dispersion enhancement structure of the curve that we derive our key results from.

These results affirm the effectiveness and reliability of our foreground subtraction approach, and demonstrate robustness against methodological variations.

\subsection{The final RM-group sample}\label{sec:finalsample}

To ensure the reliability of our RRMs and the robustness of our findings, we implemented several cuts to the data, \emph{excluding} RRMs associated with sources that meet any of the following conditions:

\begin{enumerate}
\item have linearly polarised signal-to-noise ratios $<10$, to conservatively mitigate spurious RM measurements \citep[e.g.][]{Macquart2012,Anderson2015};
\item are located at $|b|<20^\circ$ to mitigate the possible impact of small-scale RM structure in the Galactic foreground;
\item are within $10^\circ$ of the Large Magellanic Cloud, where RMs may be biased by contributions from that system (McClure-Griffiths, \emph{in prep.});
\item are situated in regions affected by imaging artifacts caused by bright, resolved radio galaxies, typically within 1-degree separation;
\item are sub-components (from source-finding) of bright, resolved radio galaxies;
\item are associated (that is, are closest in terms of $\xi$) with halos outside the mass range of interest (i.e., log10(M$_\odot$)<12.5 or log10(M$_\odot$)>14.0); or,
\item are associated with groups having fewer than three member galaxies \footnote{\citet{Tully2015} define groups with as few as a single member galaxy, explaining their rationale therein.} in the MKT catalogue;
\end{enumerate}

\begin{figure}
\centering
\includegraphics[width=0.48\textwidth]{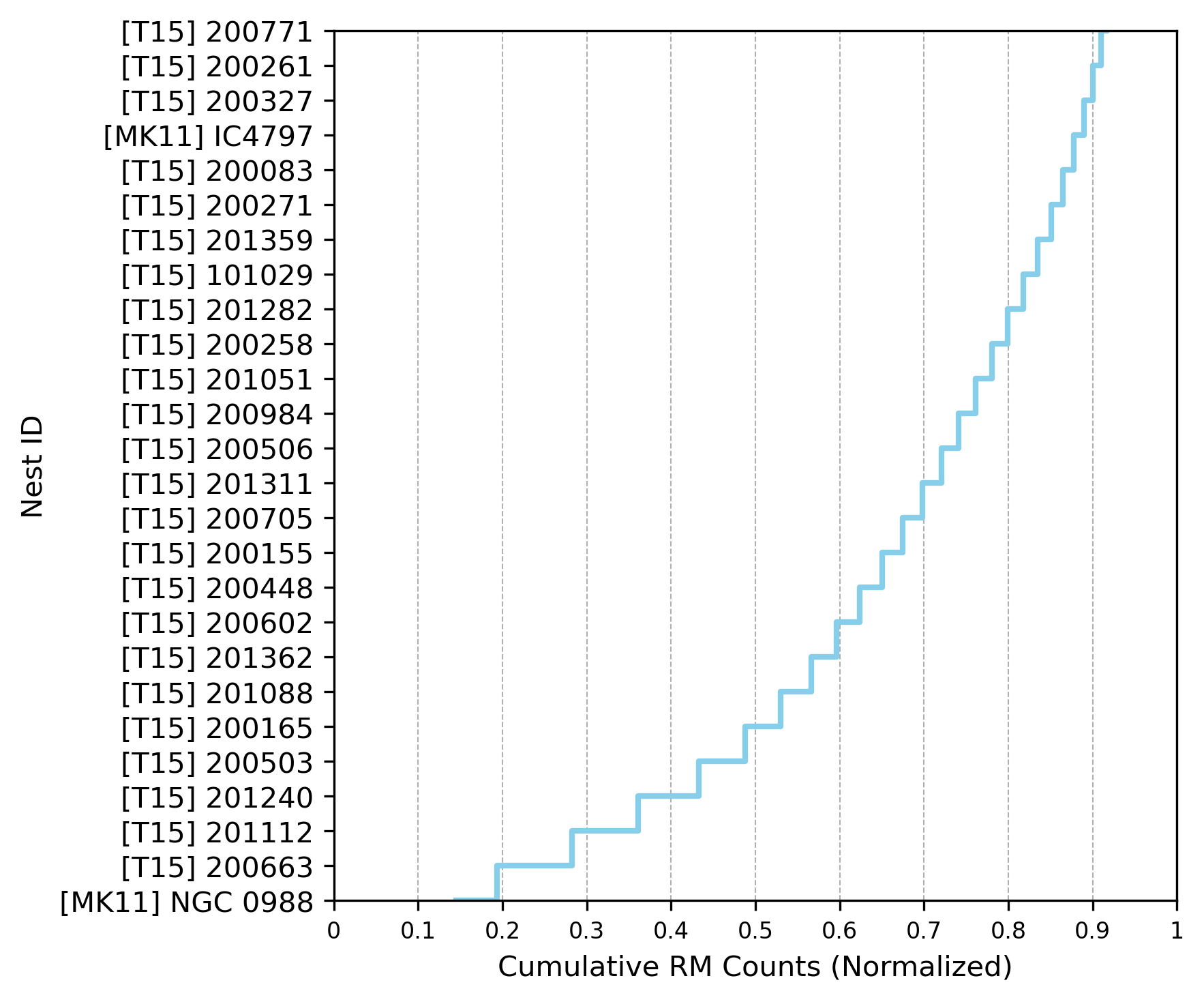}
\caption{Cumulative proportion of total RMs associated with galaxy groups in our final sample, sorted by their relative contribution to the total of 22,817 RM measurements, and truncated to include only those 26 groups contributing the first 90\% of this total. Tick marks at 0.1 intervals on the x-axis mark the cumulative proportion up to the 90\% threshold. `Nest' IDs (the IDs given to individual groups and clusters in the MKT catalogue) on the truncated y-axis identify the key contributing groups, with '[T15]' indicating that the groups (and associated group identifiers) are drawn from the \citet{Tully2015} catalogue, and '[MK11]' from the \citet{MK2011} catalogue. This plot highlights how our analysis and results are weighted towards a relatively smaller number of sources from the total sample of 55 objects.}
\label{fig:proportion_of_RMs_associated_with_groups}
\end{figure}

\begin{figure}
\centering
\includegraphics[width=0.48\textwidth]{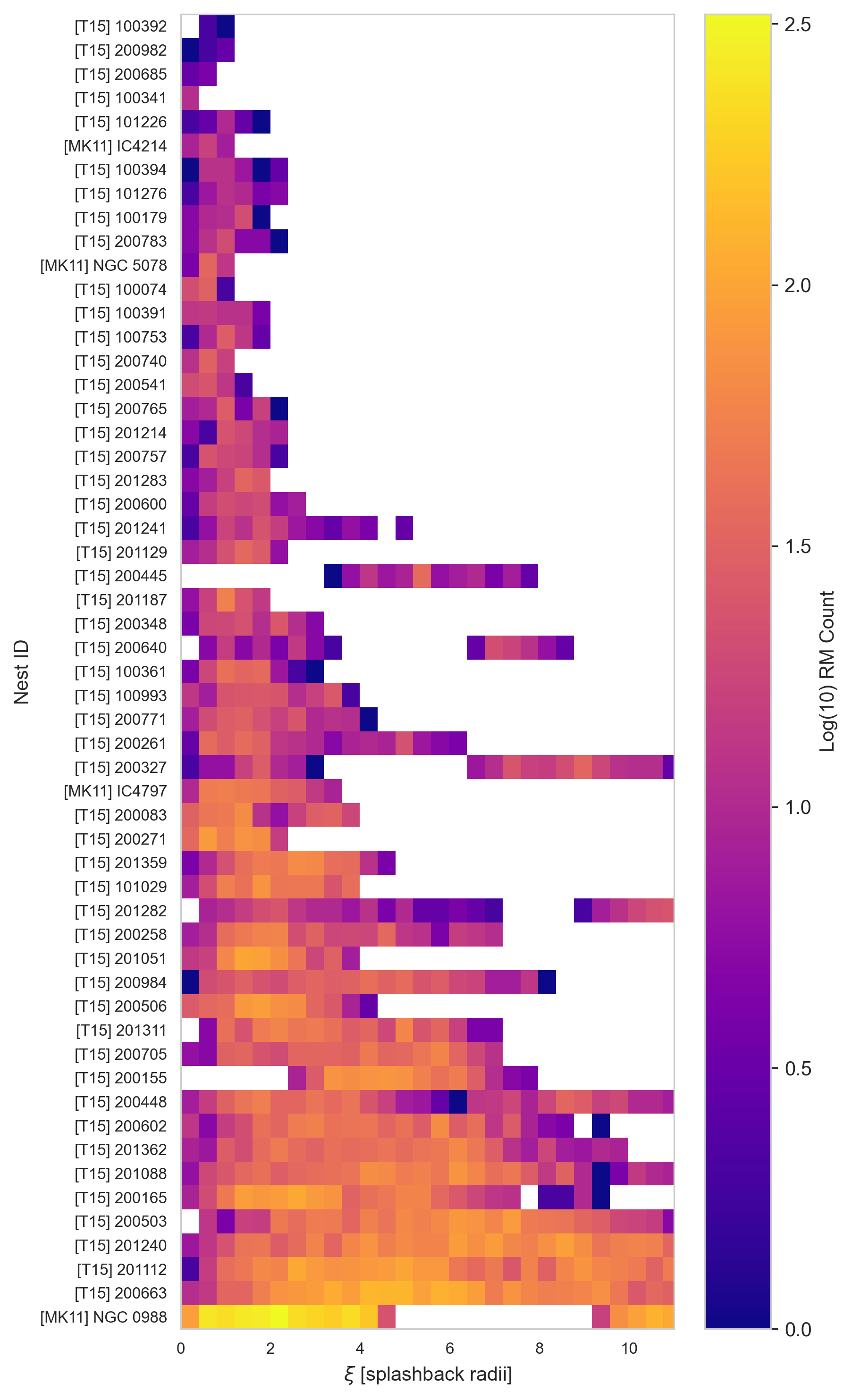}
\caption{Logarithm of RM count (color scale) binned by scaled impact parameter ($\xi$;} $x$-axis; in bins of width 0.4) and galaxy group association ($y$-axis). The latter is organised inversely according to the number of RMs associated with each group, and the Nest IDs indicate the MKT catalogue designations, mirroring the arrangement in Figure \ref{fig:proportion_of_RMs_associated_with_groups}. White areas represent ranges of $\xi$ not covered by the background RMs for each galaxy group.
\label{fig:RMs_as_function_of_groups_and_scaled_distance}
\end{figure}

Our final sample consists of 55 galaxy groups with redshifts in the range $0.005\lesssim z \lesssim0.025$, analysed using a total of 22,817 RRMs measured towards background radio sources \footnote{Radio sources brighter than $\sim$1 mJy in total intensity at $\sim$GHz frequencies --- satisfied by 98.5\% of our RM sources --- overwhelmingly lie at greater redshifts than our sample maximum of $z\approx0.025$ (e.g. \citealp{Magliocchetti2000,deZotti2010})}, covering a combined sky area of 1520 square degrees. It should be noted that the RMs are apportioned across these groups unevenly, as illustrated in Figure \ref{fig:proportion_of_RMs_associated_with_groups}. The figure shows that 90\% of the RMs are associated with half of the groups, 50\% of the RMs with only six groups, and a single group accounts for nearly 20\% of the total RMs. We further note that the distribution of RM sampling across $\xi$ ranges for galaxy groups is also uneven, as shown in Figure \ref{fig:RMs_as_function_of_groups_and_scaled_distance}. This arises from present limitations in POSSUM's areal sky coverage coupled with the tendency of nearby or massive groups to occupy larger angular scales. The situation will improve as POSSUM progresses towards full coverage of the southern sky in the coming years. We characterise the potential impact of this skewed distribution on our main findings in Section \ref{sec:Variance_by_group}.\\

The groups included in our final sample are catalogued in Table \ref{tab:groups}. The columns are Right Ascension (RA, J2000); Declination (Decl., J2000); number of group members (N Members); Splashback Radius (Mpc); Splashback Radius (Degrees);  Distance (Mpc); Halo Mass ($\times10^{13}$ M$_\odot$); Tully Nest ID; Parent SBID; Principal Galaxies Catalogue ID (PGC); Common designation of dominant galaxy (New General/Index/Messier Catalogues).

\section{Analysis and Results}\label{sec:Results}

\subsection{Residual RM vs. group-centric radius}\label{sec:RRMvsRadius}

\begin{figure}
\centering
\includegraphics[width=0.5\textwidth]{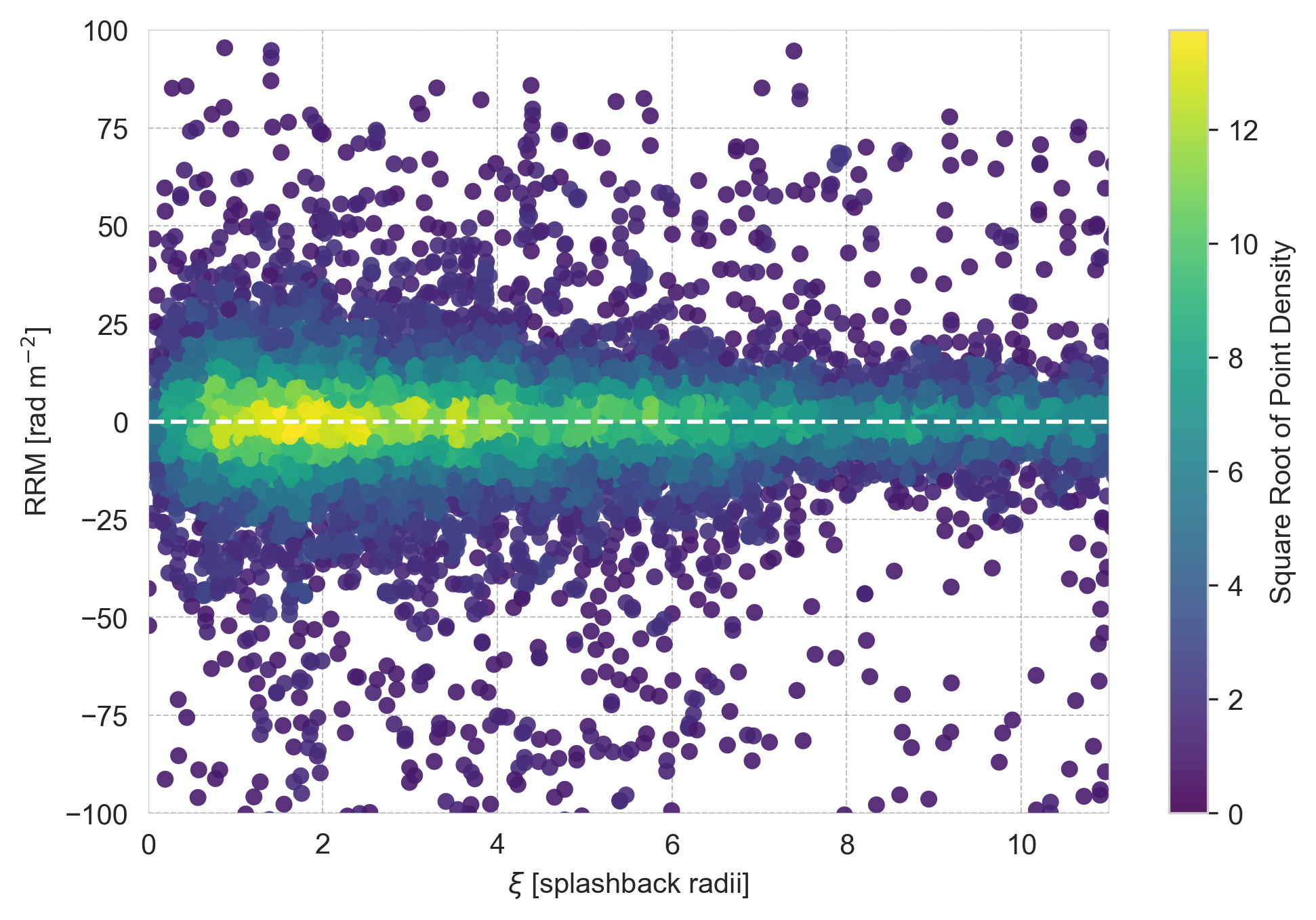}
\caption{RRMs vs. the fiducial scaled impact parameter from their nearest galaxy group centre ($\xi$), in units of splashback radii, for our sample. The points are coloured by the square root of the local point density on the plot, to better reveal structure in the distribution. The white dashed line indicates RRM$=0$\radmm.}
\label{fig:RMs_vs_fidsep_density_new}
\end{figure}

In Figure \ref{fig:RMs_vs_fidsep_density_new}, we present the distribution of RRMs plotted against multiples of the splashback radius. It is first worth explaining why there are fewer sources at large splashback radii, when the opposite might be expected. This is mainly due to the proximity of groups to one another and our sky coverage. That is, it becomes increasingly difficult to find regions of POSSUM survey coverage that are far from the nearest group without encountering other groups, especially with some nearby massive groups having large angular radii.

\begin{figure}
\centering
\includegraphics[width=0.48\textwidth]{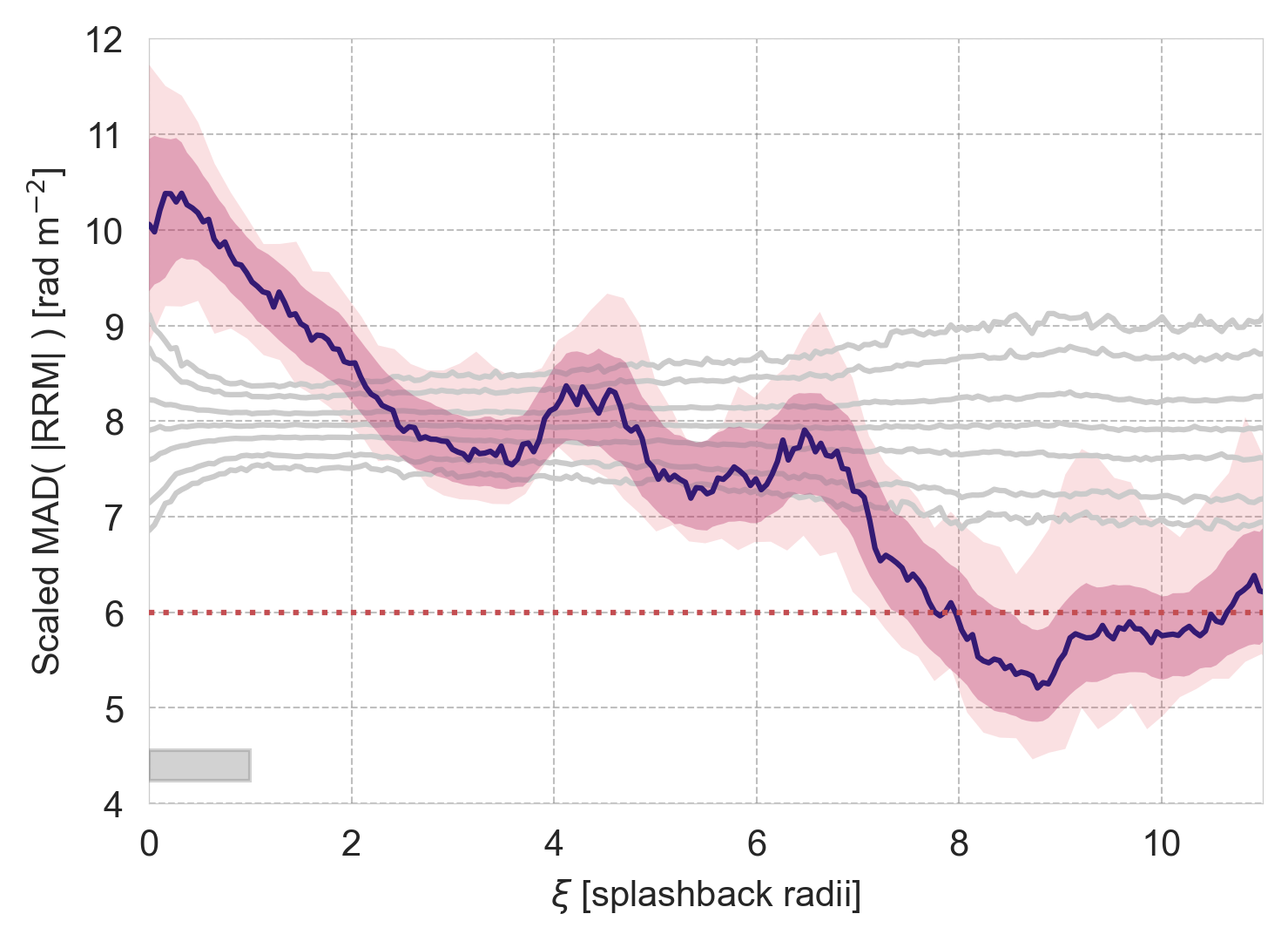}
\caption{Analysis of RRM scatter as a function of the fiducial scaled impact parameter $\xi$. The purple line is the calculated $\sigma_{\text{RRM,corr}}$ value within a sliding window, the width of which is indicated by the grey-shaded bar at lower left. Surrounding the purple line, the dark magenta shaded region delineates the 95\% confidence interval on the statistic, obtained through bootstrap resampling. The light magenta shaded region delineates the 95\% confidence interval on the statistic under the action of bootstrap resampling by group association, described in Section \ref{sec:Variance_by_group}. The red dotted line, set at $\sigma_\text{RRM}=6$ \radmm, shows the expected standard deviation of RRM values for extragalactic point-sources as per \citet{Taylor2024, Schnitzeler2010}. The light gray lines show the percentile distribution (1st/99th, 5th/95th, 25th/75th, and 50th) of 1000 $\sigma_\text{RRM}$ curves generated identically to the main curve, but derived from randomly re-pairing RRM values with source positions in each case, illustrating the behaviour of the metric when the spatial relationship between source position, RRM, and group position is broken.}
\label{fig:sliding_window_MAD_plus_control}
\end{figure}

Visual inspection of the figure suggests that RRMs penetrating group halos at smaller multiples of the splashback radius show greater dispersion than those at larger multiples. To quantify this, we calculated the Median Absolute Deviation (MAD) within a sliding window of width $\Delta \xi=1$ across the entire range of $\xi$. Multiplying the MAD by 1.4826 provides a robust estimate of the standard deviation \citep{RC1993} of the RRMs, denoted as $\sigma_\text{RRM}$. The total uncertainty on our RRMs is typically less than a few \radmm (Section \ref{sec:RM foreground}), and so does not drive the scatter seen in Figure \ref{fig:RMs_vs_fidsep_density_new}. Nevertheless, we adopt a method to remove the small impact of RRM uncertainties following \citet{Dolag2001}. We compute the corrected RM scatter ($\sigma_{\text{RM,cor}}$) by subtracting the mean contribution of measurement errors from the observed scatter in quadrature according to:

\begin{equation}
\sigma_{\text{RRM,corr}} = \sqrt{\left(1.4826 \times \text{MAD}\right)^2 - \frac{\sum_{i=1}^{N} (\delta \text{RRM}_i)^2}{N - 1}}
\end{equation}

\noindent where $\delta \text{RRM}_i$ is the total uncertainty on the RRM of an individual data point $i$ in the sliding window, and $N$ is the number of data points in the sliding window. We determined 95\% confidence intervals for this statistic using $10^4$ bootstrap resamples within the sliding window. The results are shown in Figure \ref{fig:sliding_window_MAD_plus_control}.

The expected RM dispersion for extragalactic point-like field sources, as per \citet{Schnitzeler2010, Taylor2024}, is $\sigma_\text{RRM}\approx 6$ \radmm, which falls within the 95\% confidence interval of the MAD(RRM) curve for all $\xi>7$ (Figure \ref{fig:sliding_window_MAD_plus_control}), where the mean measured value is $\sigma_\text{RRM}=7.1$ \radmm. This similarity suggests that the RRM contribution from gas associated with galaxy groups is minimal beyond seven splashback radii, allowing us to categorise these as 'field RMs'. Conversely, inside $\xi\sim7$, the RRM dispersion significantly exceeds $\sigma_\text{RRM}=6$ \radmm.

To evaluate the statistical significance of the observed differences in |RRM| distributions between field sources and non-field sources (those inside $\xi=7$), we employed the Kruskal-Wallis H-test, which assesses differences in median values, and the two-sample Kolmogorov-Smirnov test, which examines the overall distributions. The results, detailed in Table \ref{tab:stats_results_conformatory_test}, strongly support the rejection of the null hypothesis\\

\begin{table}
\centering
\caption{Summary of Statistical Test Results comparing the magnitudes of field RRMs to non-field RRMs (defined by $\xi>7$ and $\xi<7$, respectively). Results are quoted to 3 significant figures.}
\begin{tabular}{lcc}
\hline
Test Name & Test Statistic & p-value \\ \hline
Kruskal-Wallis H-test & 141 & $1.48 \times 10^{-32}$ \\
Kolmogorov-Smirnov Test & 0.0973 & $2.48 \times 10^{-33}$ \\
\hline
\end{tabular}
\label{tab:stats_results_conformatory_test}
\end{table}

Visual inspection of the MAD(RRM) curve identifies two distinct increases in value as it nears group centers from larger scaled distances: first at $\xi\sim7$ and then at $\xi\sim2$. On this basis, we change our approach from hypothesis testing---that is, whether RRM dispersion in non-field sources is higher compared to field sources---to an exploratory analysis. We apply the same statistical tests to compare distributions of absolute RRMs bounded by these $\xi$ ranges identified visually. This analysis assesses whether the observed differences could reasonably arise by chance.

The results, shown in Table \ref{tab:stats_comparison_splashback}, firmly reject the null hypothesis across all comparisons. The magnitude of the evidence is such that systematic influences like imperfect Galactic foreground RM subtraction and cosmic variance are likely to be more significant than raw statistical variations. These challenges represent key areas of ongoing research in POSSUM (e.g. \citealp{Khadir2024inprep}). Nonetheless, we contend that the statistical evidence supports the initial visual findings of distinct steps in the MAD(RRM) curve, confirming substantial differences in the |RRM| distributions between these ranges.

\begin{table}
\centering
\caption{Statistical comparison of |RRM| distributions across different ranges of $\xi$. Results are quoted to 3 significant figures.}
\begin{tabular}{lcc}
\hline
Comparison & Statistic & p-value \\ \hline
Kruskal ($\xi<2$ vs. $2\lesssim \xi<7$) & 40.4 & $2.04 \times 10^{-10}$ \\
Kruskal ($\xi<2$ vs. $\xi>7$) & 179 & $9.61 \times 10^{-41}$ \\
Kruskal ($2\lesssim \xi<7$ vs. $\xi>7$) & 82.9 & $8.49 \times 10^{-20}$ \\
KS ($\xi<2$ vs. $2\lesssim \xi<7$) & 0.053 & $3.05 \times 10^{-10}$ \\
KS ($\xi<2$ vs. $\xi>7$) & 0.124 & $7.49 \times 10^{-39}$ \\
KS ($2\lesssim \xi<7$ vs. $\xi>7$) & 0.082 & $2.17 \times 10^{-21}$ \\
\hline
\end{tabular}
\label{tab:stats_comparison_splashback}
\end{table}

\subsection{Null test to confirm RRM-group association}\label{sec:Nulltest}

To verify that the observed correlation between RRM dispersion enhancements and group-relative positions was not coincidental, we generated 1,000 MAD(RRM) curves from RRM values that were randomly shuffled relative to their sky positions, maintaining the original distribution of RRMs but breaking the spatial link to group positions. The 1st/99th, 5th/95th, 25th/75th, and 50th percentiles of these curves, calculated as a function of fiducial radius, are depicted as gray lines in Figure \ref{fig:sliding_window_MAD_plus_control}.

The MAD(RRM) curves generated from randomly reshuffled RRMs exhibit no significant trend with fiducial radius, consistently showing a narrow range of RRM dispersion values centered around 8.0 \radmm. In contrast, the actual MAD(RRM) curve, including its 95\% confidence intervals, distinctly surpasses the 1st/99th percentiles at $0\leq \xi<2$ and falls below at $\xi>7$. This difference provides robust evidence linking RRM dispersion to group-centric positions and thus gas associated with galaxy groups.

\subsection{RRM enhancement is not driven by the circumgalactic medium of group member galaxies}\label{sec:notCGM}

In this section, we consider whether our results could be attributed to the circumgalactic medium of group member galaxies, which is magnetised and may contribute measurable RM excesses of up to a few \radmm (e.g. \citealp{Heesen2023,Jung2023,Ramesh2023,Pakmor2024}).

The scalings of mass and splashback radius with K-band luminosity described in Section \ref{sec:groupoffsets} apply to objects with K-band luminosities down to $10^{10}~L_\odot$, typical of mid-sized spiral galaxies \citep{Tully2015}. In our sample, over 95\% of MKT group member galaxies have K-band luminosities above $10^{10}~L_\odot$. Thus, we matched each RM with its closest group galaxy, determining its fiducial radius in terms of multiples of the splashback radius, in the same way as we did for galaxy groups themselves.

Of the RMs within the splashback radius of a galaxy group, only 78 (or 3\%) also intersect the splashback radius of a member galaxy in the galaxy group. Upon excluding these RMs and regenerating the plots in Figures \ref{fig:RMs_vs_fidsep_density_new} and \ref{fig:sliding_window_MAD_plus_control}, along with the statistics in Tables \ref{tab:stats_results_conformatory_test} and \ref{tab:stats_comparison_splashback}, the differences were negligible. We conclude that the observed increase in RRM dispersion is not driven in this case by the CGM of individual group member galaxies.

\subsection{Variability in RRM Enhancement Results by Group Selection}\label{sec:Variance_by_group}

We leave a comprehensive examination of RRM enhancements categorised by group properties to future work. Nonetheless, it is crucial to determine whether the RRM dispersion signal is primarily linked to a few individual groups or is representative of a significant portion of the sample. The former is plausible, because our RMs are not evenly apportioned among the galaxy groups in our sample, nor do they consistently probe the entire range of $\xi$ for each group (see Figures \ref{fig:proportion_of_RMs_associated_with_groups} and \ref{fig:RMs_as_function_of_groups_and_scaled_distance}, and Section \ref{sec:finalsample}). This means that specific groups could have an outsized effect on the RRM dispersion over particular ranges of $\xi$.

To address this, we conducted a standard bootstrap resampling experiment, a widely used method to assess the variance of a sample in statistical analysis. In a bootstrap experiment, sampling with replacement is performed from the full dataset to create new samples of the same size as the original dataset. This allows for an estimation of the underlying sample variance, as some objects may be repeated while others may be omitted in each sample.

In this case, we randomly selected 55 objects from our full sample of galaxy groups, with replacement, 1,000 times. For each of these resampled datasets, we recalculated the MAD(RRM) curve for the RRMs associated with these selections. From these recalculated curves, we derived new 95\% confidence intervals for the true MAD(RRM) curve. This estimates the variance of the RRMs with respect to the group sample. For example, if the variance structure of the true MAD(RRM) curve is driven by only a handful of groups, the bootstrap resamples are less likely to consistently include these groups, resulting in larger confidence intervals centered around a value closer to 6 \radmm.

The newly calculated confidence intervals are depicted as the light-pink shaded region in Figure \ref{fig:sliding_window_MAD_plus_control}. These intervals demonstrate that the core characteristics of the true MAD(RRM) curve are preserved under this resampling, although the 95\% confidence interval becomes approximately 60\% wider than that obtained through the RM bootstrapping (dark-magenta shaded region in Figure \ref{fig:sliding_window_MAD_plus_control}).

The results show that while the specific composition of our galaxy group sample introduces some variance to the RRM dispersion, this effect is minor and does not significantly alter the principal conclusions of our study.

It is also clear that the 95\% confidence interval's lower bound for $0<\xi<2$ lies well above the mean value of the true MAD(RRM) curve for $2<\xi<7$. This implies a significant increase in RRM dispersion at $\xi<2$ among most of our galaxy groups, in turn implying that most galaxy host a detectable, Faraday-rotating IGrM phase. We plan to further quantify and constrain this observation with a larger group and RM sample generated by POSSUM in future work.

For regions $2<\xi<7$ and $\xi>7$, the MAD(RRM) curves and their 95\% confidence intervals are also inconsistent, indicating that our group sample also demonstrate elevated RRM dispersion well beyond their splashback radius, compared to unresolved radio galaxies whose lines of sight do not intersect with group halos and their associated gas at all.

\subsection{Magnitude of RRM enhancements}\label{sec:RM magnitude}

We propose that the enhancements in RRM dispersion detailed in Section \ref{sec:RRMvsRadius} delineate three distinct regions of magnetised gas: (1) within groups where $\xi<2$; (2) external to groups for $2<\xi<7$; and (3) beyond $\xi>7$, where RRM dispersion is consistent with intrinsic contributions from unresolved extragalactic radio sources, as discussed by \citet{Schnitzeler2010} and \citet{Taylor2024}. The magnitude of RRM contributions can be calculated for each region by subtracting the relevant $\sigma_\text{RRM}$ values in quadrature. Specifically, we use the mean value from the MAD curve for the outer regions and the peak value for the inner group regions (since the curve forms a plateau in the former region, and rises to a peak in the latter, respectively, as shown in Figure \ref{fig:sliding_window_MAD_plus_control}). Consequently, and with the caveat that the outer regions are probed by successively smaller fractions of our total group sample (see Figure \ref{fig:RMs_as_function_of_groups_and_scaled_distance}), the RRM dispersion contributed by gas at $2<\xi<7$ is calculated to be $\sigma_\text{RRM, 2<As<7} = 4.2 \pm 1.2$ \radmm, and by gas within $\xi<2$, it is $6.9\pm1.8$ \radmm, where the uncertainties have been obtained by appropriately modifying the bootstrap confidence intervals calculated in Section \ref{sec:Variance_by_group}.

\section{Discussion}\label{sec:discussion}

Our findings confirm the presence of a magnetised, ionised IGrM within most galaxy groups, detectable up to twice the splashback radius. Additionally, we identified a second, more extended component of gas that extends out to seven splashback radii. The associated enhancements in RRM dispersion ($6.9 \pm 1.8$ \radmm and $4.2 \pm 1.2$ \radmm) are consistent with the drop in expected RM contributions from $\sim10$ \radmm within dense group environments to $\sim1$ \radmm within sparser cosmic filaments, as modeled by \citet{Akahori2010}. This distinct transition mirrors findings by \citet{Gouin2022}, who observed a shift from predominantly hot, virialised gas in core regions of groups to a more tenuous WHIM gas extending well beyond their virial boundaries (also see \citealp{Oppenheimer2021, Anderson2021}). Indeed, these two gas phases---the IGrM and the surrounding WHIM---are the only plausible candidates for providing the enhanced RRM dispersions that we find in Section \ref{sec:Results}, which we discuss in more detail below. Broadly though, our findings underscore that Faraday RM grids serve as an effective tracer of magnetised gas in circum-halo environments.

\subsection{The Intragroup Medium}\label{sec:firsthump}

\subsubsection{Prevalence and detectability}\label{sec:hotIGrMPrevalence}

We are hesitant to quantify the fraction of galaxy groups that host a magnetised, ionised IGrM, owing to the present limited sky coverage of POSSUM data and associated potential biases (Section \ref{sec:finalsample}). Nevertheless, a large fraction of our sample groups contribute to RRM dispersion (Section \ref{sec:Variance_by_group}), so most galaxy groups must contain such a gas phase. This could represent the hot, volume-filling gas phase implied by UV absorption IGrM studies \citep{Stocke2019, McCabe2021}. The detection rate of this gas via Faraday rotation likely surpasses the 10–20\% of compact groups harboring relativistic magnetised gas detected via radio synchrotron emission \citep{NW2019}, and notably, the $\sim30$\% of groups/clusters whose ICM is detected via current X-ray observations (e.g. Appendix C of \citealp{Xu2022}). This may indicate that the gas illuminated by RMs is too cool or sparse to be detected in X-rays, perhaps because the gas has not reached the virial temperature in accreting halos (e.g. \citealp{Pratt2021}), or because magnetic fields can provide pressure support that lowers the virial temperature \citep{Lovisari2021}. S-Z observations will eventually provide another basis for comparison \citep{Pandey2023}, but is not yet generally capable of detecting gas in group-mass halos.

\subsubsection{Magnetisation}\label{sec:hotIGrMMagnetisation}

Magnetic fields could play a pivotal role in shaping astrophysical processes within galaxy groups, from modulating energy dispersal from black hole jets to affecting pre-processing in galactic evolution \citep[e.g.][and references therein]{Oppenheimer2021}. Evidence of magnetised group gas has been increasingly observed through Faraday rotation, including in the thermal IGrM of the Antennae system \citep{Basu2017}, the Magellanic Bridge \citep{Kaczmarek2017}, Stephan's Quintet \citep{NW2020}, and in the (distinct) synchrotron-emitting relativistic gas phase in about 10-20\% of compact groups \citep{NW2019}. Notably, the estimated field strengths suggest that magnetic fields could be dynamically important in these contexts.

Presently, RM grid-based estimates of magnetic field strengths in cluster environments carry fractional uncertainties of hundreds of per cent \citep{Johnson2020}, even when the thermal particle density is well-known, which is not the case here. Despite these uncertainties, coarse estimates of the plausible range of magnetic field strengths required to produce the observed RM dispersions remain valuable for testing the validity of our results, as well as in their own right. Thus, following \citet{Gaensler2001}, we have the relation:
\begin{equation}
\label{eqn:BfromSigRM}
\frac{ \sigma_{\text{RM}} } { \text{rad/m}^2 } = \frac{ 812 }{2 \sqrt{3} }
\left( \frac{ n_e }{ \text{cm}^{-3} } \right)
\left( \frac{ B }{ \mu\text{G} } \right)
\sqrt{ \left( \frac{ L }{ \text{kpc} } \right)
\left( \frac{ l }{ \text{kpc} } \right) },
\end{equation}

where $\sigma_{\text{RM}}$ is the RM dispersion, $n_e$ the electron density, $B$ the magnetic field strength, $L$ the IGrM path length, and $l$ the magnetic field correlation scale. 

A second quantity of interest is the plasma beta defined as:

\begin{equation}
\beta = \frac{P_{\text{thermal}}}{P_{\text{magnetic}}} = \frac{nk_BT}{ B^2 \slash 8\pi}
\end{equation}

\noindent where in turn $P_{\text{thermal}} = n_ek_BT$ is the thermal pressure, $k_B$ is the Boltzmann constant, $T$ is the temperature of the plasma, $P_{\text{magnetic}} = B^2 \slash 8\pi$ is the magnetic pressure (in CGS units), and all other quantities are as previously defined. A plasma beta greater than one ($\beta > 1$) indicates that thermal pressure dominates over magnetic pressure, whereas a plasma beta less than one ($\beta < 1$) implies that magnetic pressure is dominant.

Using Monte Carlo methods, we estimated $B$ as well as $\beta$ for the IGrM  of our galaxy group sample by randomly selecting $\sigma_{\text{RM}}$ values from a range of [9.25, 11.5] \radmm, representing the percentile range at the peak of the curve in Figure \ref{fig:sliding_window_MAD_plus_control}, and then adjusted by subtracting the mean $\sigma_{\text{RM}}$ enhancement values for $2<\xi<7$ and $\xi>7$ of 4.2 \radmm and 6 \radmm respectively in quadrature. The electron density $n_e$ was selected from [$10^{-5}$, $10^{-4}$] cm$^{-3}$ (e.g. \citealp{Eckert2013,Nugent2020,Angelinelli2022,RD2023}), the magnetic field correlation scale $l$ was selected from [$1/100$,1] times the splashback diameter $L$, which itself was randomly selected from our galaxy group sample. Finally, $T$ was selected from [$10^{6}$, $10^{8}$] K (e.g. see Figure 1 of \citealp{Gouin2022}).

\begin{figure}
\centering
\includegraphics[width=0.48\textwidth]{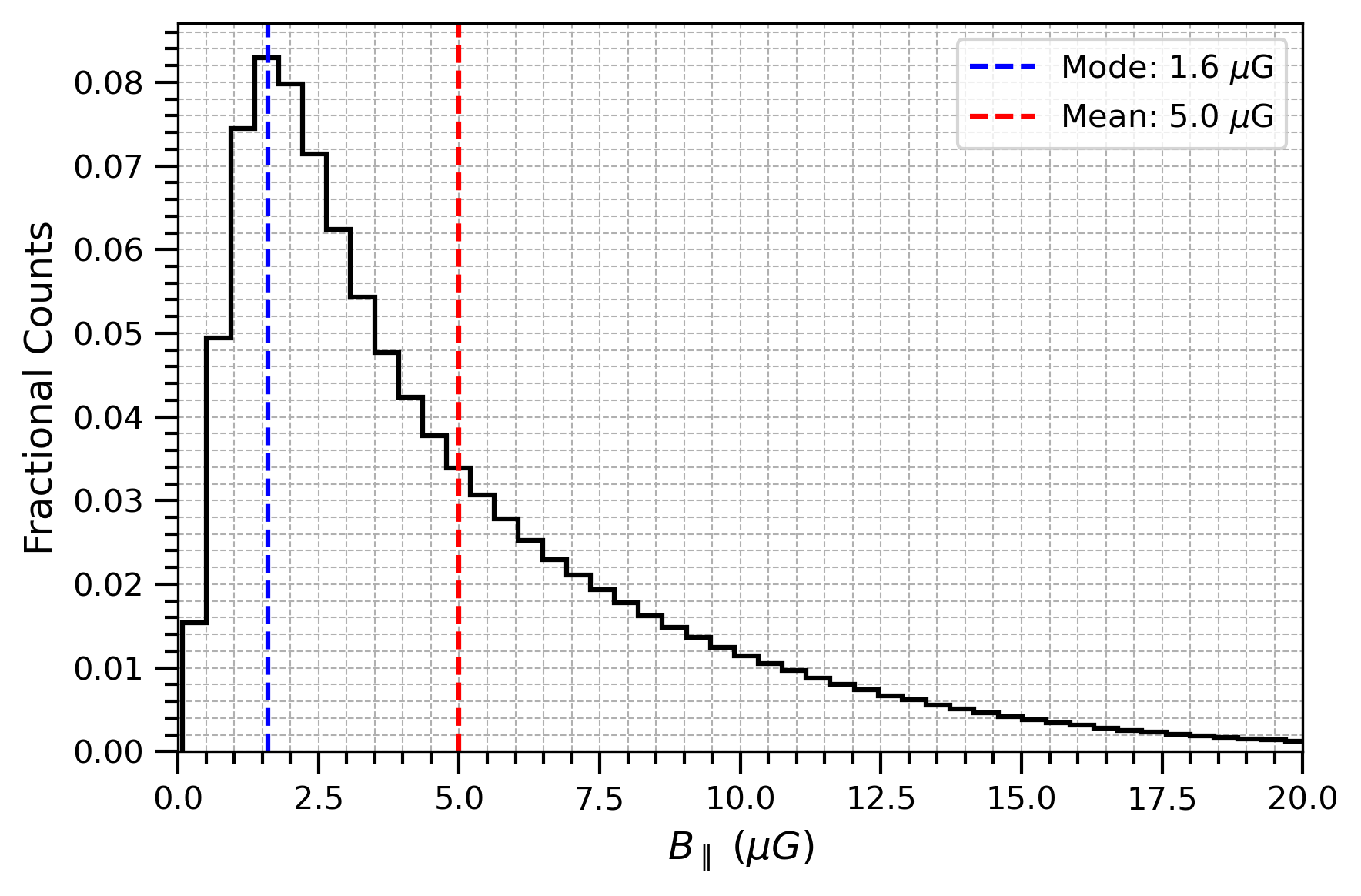}
\caption{Histogram of $B$ values described in Section \ref{sec:hotIGrMMagnetisation}.}
\label{fig:B_field_estimates}
\end{figure}

Randomly sampling these prior distributions $10^7$ times and solving for $B$, we obtain the distribution shown in Figure \ref{fig:B_field_estimates}. The mode and mean of the $B$ distribution is 1.6 $\muup$G and 5.0 $\muup$G, respectively, with a long tail extending to $\sim20\muup$G. The magnetic reversal scale, a critical yet less certain parameter, affects these estimations quite strongly. Restricting the number of line-of-sight reversals to a maximum of 10 alters the mode and mean to 0.6 $\muup$G and 1.7 $\muup$G, respectively. In any case, the total magnetic field strength broadly agrees with estimates cited in the IGrM studies listed above---typically $\sim$ a few $\muup$G. 

Our estimate of the plasma $\beta$ comes with several caveats and should be considered indicative only. Astrophysically, even in single galaxy groups, the $\beta$ value is likely sensitive to local conditions and varies significantly throughout the group environment. However, our derivation is based on a stacking experiment that encompasses a broad range of masses, averaged out to twice the splashback radius, and does not account for potential astrophysical correlations among the parameters involved. Moreover, these values are drawn from ranges selected to be only broadly representative of group environments. Despite these limitations, our Monte Carlo sampling yields a median $\beta$ value of 0.94, with 10th and 90th percentile values of 0.03 and 9.8, respectively. The $\beta$ value exceeds 1 in 49\% of the samples.

These values are generally smaller (i.e., more magnetically dominated) than is typical in massive galaxy cluster ICMs (where $\beta$ is more typically 10--100; e.g., \citealp{vazza2021}), driven primarily by the lower electron densities and temperatures in our groups compared to clusters, while still supporting the microgauss-level magnetic field strengths cited above. This, in turn, may be related to the fundamentally lesser influence of gravity compared to other baryonic processes at work in groups, which we discuss in more detail in Section \ref{sec:hotGasExtent} below. It may also suggest that RM enhancements are efficient tracers of low-$\beta$ plasma regions and processes, potentially biasing the value we derive as a result. Regardless, the fact that the percentile values straddle a ratio of 1 suggest that magnetic fields may play a significant role in the dynamics of the IGrM.

\subsection{The Cosmic Web}\label{sec:secondhump}

\subsubsection{Extent of magnetised thermal gas}\label{sec:hotGasExtent}

Our observations reveal that the magnetised thermal gas associated with galaxy groups extends well beyond their splashback radii---that is, their effective gravitational boundaries---with significant \sigrrm enhancements persisting up to seven splashback radii before tapering off to the expected background level of $\sim6$ \radmm. Such gas must be in a dynamic rather than bound state, likely being accreted from the surrounding cosmic web or expelled towards it. 

Other findings in the recent literature also hint at the presence of gas at the periphery of group- and cluster-mass halos. \citet{Anderson2021} used pilot POSSUM data to show that RM dispersion enhancements illuminated cosmic gas substantially further out in the Fornax cluster than existing X-ray observations, towards the virial radius, and additionally highlighted features of the gas related to the dynamics of the cluster including shocks and a swept-back tails induced by merging substructures. Deviations from self-similar scaling relations at the group scale, notably the reduced gas fractions, further underline the dynamic nature of group environments, and suggest that gas is being systematically expelled from group centres \citep{Angelinelli2022, Ayromlou2023, RR2023}. UV absorption studies have detected ionised gas far beyond the virial radii of galaxy groups \citep{Stocke2019, McCabe2021}, and new groups and clusters with extended, flattened X-ray profiles have been reported, suggesting the presence of significant ionised gas reservoirs in halo outskirts \citep{Xu2022}. Similarly, the detection of magnetised `megahalos' around cluster systems by \citet{Cuciti2022}, although involving more massive systems than considered here, suggests that magnetised gas at halo boundaries might be more widespread than anticipated. 

With respect to our results specifically---that is, the RRM dispersion enhancement extending from 2 to 7 splashback radii (or $\sim$ 3 to 7 Mpc) that we attribute to the WHIM---we point out that galaxy groups are thought to be embedded in cosmic filaments (of typical width $\sim3$ Mpc; e.g. \citealp{OSullivan2019,Carretti2022}) and magnetised bubbles that span between 1 and 10 Mpc in diameter; \citealp{AG2021}). Simulations further suggest that the cosmic budget of `missing baryons' becomes closed at fiducial separations of between and few and 10 times the splashback radius after being ejected by AGN activity \citep{Ayromlou2023}. Thus, we claim to have detected such material directly, as other recent studies have also focused on achieving with LOFAR RM data \citep{BN2024,Bondarenko2024}. Future more detailed comparisons of POSSUM and LOFAR data with simulations will be fruitful to understand the relative strengths of each dataset, and to help constrain baryonic feedback and flow models in simulations. 

\subsubsection{Magnetisation}\label{sec:hotGasB}

We can repeat the analysis of Section \ref{sec:hotIGrMMagnetisation}, but here substitute numbers appropriate for the cosmic web WHIM. Specifically, we select $\sigma_{\text{RM}}$ values from a range of [7.0, 9.0] \radmm, representing the percentile range of the plateau in Figure \ref{fig:sliding_window_MAD_plus_control}, and then adjusted by subtracting the expected background RM dispersion of 6 \radmm in quadrature. The electron density $n_e$ was selected from [$10^{-5.5}$, $10^{-4.5}$] cm$^{-3}$ (e.g. \citealp{Akahori2010,AR2011,OSullivan2019,Carretti2022}), the total filament width $L$ was taken to be seven times the splashback diameter of the groups in our sample, which was randomly selected for each Monte Carlo iteration, and the magnetic field reversal scale $l$ was selected from [0.1,1] times the filament width, reflecting the relatively more ordered fields expected in WHIM filaments \citep{Akahori2010}. Finally, $T$ was selected from [$10^{5.5}$, $10^{7}$] K (see Figure 1 of \citealp{Gouin2022}), and we assume the plasma is fully ionised.

\begin{figure}
\centering
\includegraphics[width=0.48\textwidth]{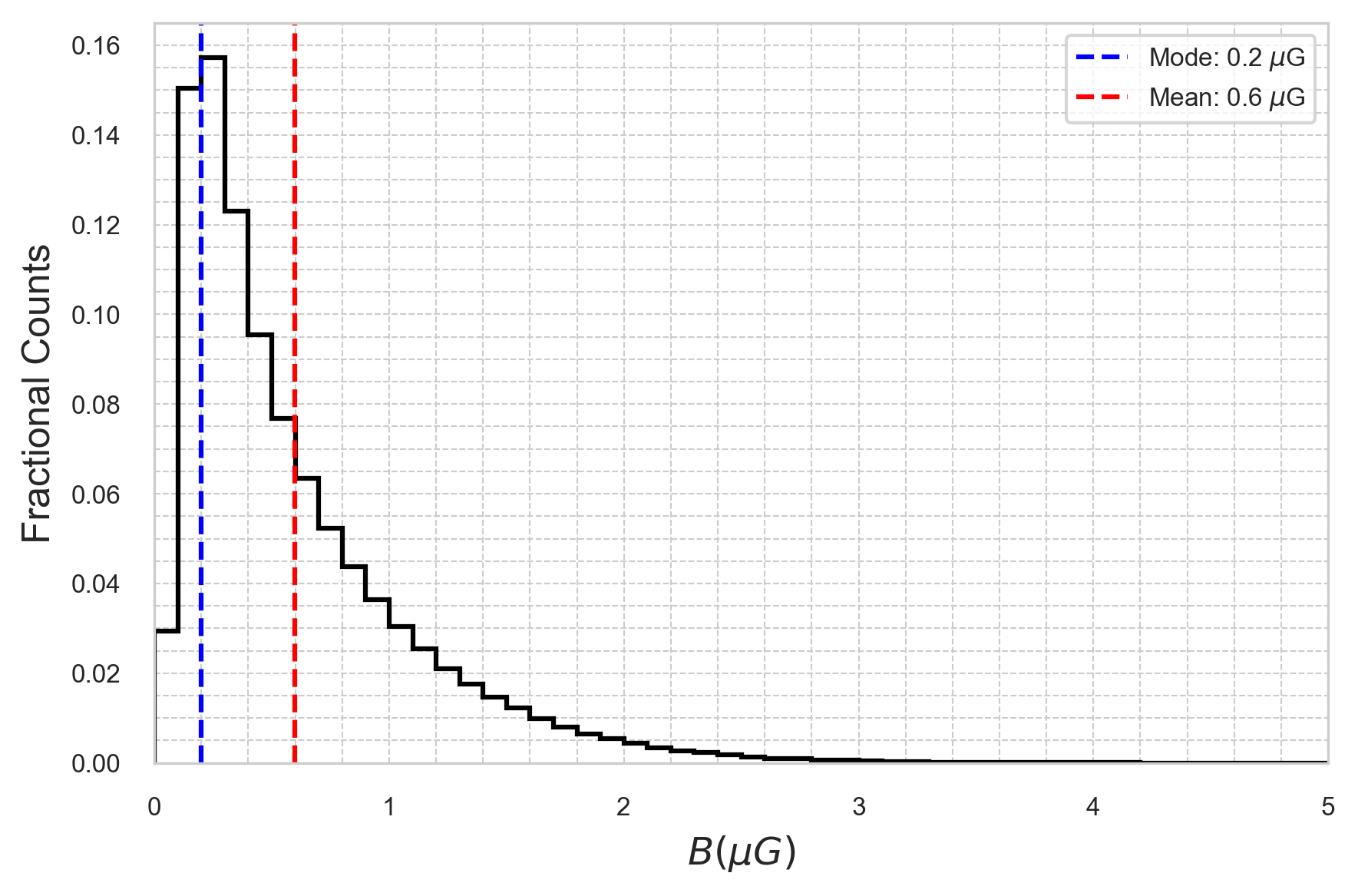}
\caption{Histogram of $B$ values described in Section \ref{sec:hotGasB}.}
\label{fig:B_field_estimates_WHIM}
\end{figure}

The resulting distribution for $B$ is shown in Figure \ref{fig:B_field_estimates_WHIM}. In this case, the mode and mean of the $B$ distribution is 0.2 $\muup$G and 0.6 $\muup$G, respectively, with a tail extending to $\sim3\muup$G. Constraining the number of line-of-sight reversals has relatively little impact here. These values are at the upper end of predictions for individual cosmic web filaments by e.g. \citet{Akahori2010}, and about an order-of-magnitude higher than the more typical predicted value of 10 nG, and initial RM-based measurements of $\sim30$ nG \citep{Carretti2022}, in such environments. We do not see these results as being in tension: The \citet{Carretti2023} observations mainly probe regions between 3 and 20 splashback radii \footnote{This assumes a conservative conversion here of $r_{\text{splashback}}\approx2\times r_{200c}$ \citep{More2015} --- the standard fiducial cluster boundary chosen by \cite{Carretti2022}} from massive clusters by design, which is a very different environment from the group-centric environment that we probe here. That is, we believe we are probing denser environments in the WHIM that build up around groups, whether by accretion or outflow.  

Under our Monte Carlo sampling, the plasma $\beta$ (again with the caveats cited in Section \ref{sec:hotIGrMMagnetisation}) yields a median value of 3.1, with 10th and 90th percentile values of 0.2 and 26.5, respectively. $\beta$ exceeds 1 in 68\% of the samples, but the fact that a significant fraction of values do not indicates that the magnetic fields may be strong enough to be dynamically important in this context. For comparison, the recent detection of magnetic fields of strength $\sim10$--84 nG in cosmic web filaments with overdensities ranging from $\sim10$--60 \citep{Carretti2023} produces $\beta$ values (we estimate here) of $\sim150$ down to $\sim10$. Thus, while the plasma $\beta$ in the region of the cosmic web near galaxy groups probed in this work overlaps with that of the sparser cosmic web filaments probed by \citet{Carretti2023}, it appears to be more magnetically dominated on average. We speculate this could be the result of AGN expelling magnetised gas into the region around the groups, as we also invoked to explain the extent of this gas above.

\subsection{RM Grids as a Prime Observational Tracer of Circum-Halo environments}\label{sec:RM_bridges}

Our results underscore the effectiveness of using cosmic magnetic fields as in-situ probes to reveal the presence, structure, and dynamics of cosmic gas in the vicinity of galaxy groups, building on work by e.g. \citet{Anderson2021} in the poor Fornax cluster. This highlights the value of RM analysis in this specific context of uncovering new aspects of the magnetised thermal IGrM and beyond, particularly with the dense RM grid of the POSSUM survey, and future RM grids of the SKA era \citep{JH2015,Heald2020}. 

RM grids thus appear to be a promising and relatively untapped resource for hunting and studying gas at halo peripheries, adding to their long-established value in the context of the canonical ICM present in massive cluster cores (e.g. \citealp{Clarke2001}). Possible high-value targets in these locations include accretion shocks and streams, merger shocks, and links between the IGrM/ICM and WHIM gas in the broader cosmic web. 

Since Rotation Measures (RMs) depend on electron density, magnetic field strength, and path length (refer to Equation \ref{eqn:RM}), integrating RM data with observations from X-ray emissions (e.g. \citealp{Bulbul2024}), the Sunyaev-Zel'dovich Effect (e.g. \citealp{Pandey2023}), and now Fast Radio Burst Dispersion Measures \citep{CR2022,Medlock2021,RR2024,Medlock2024,Theis2024} will eventually enable the measurement of the strength and orientation of magnetic fields associated with galaxy group gas. Given that magnetic fields typically dragged along with plasma flows in astrophysical environments (that is, when plasma $\beta > 1$), they naturally serve as in situ probes for detecting the presence, flow characteristics, and behavior of intergalactic gas. This approach can be likened to conducting a cosmic magnetic "dye tracer" experiment, albeit in a limited and metaphorical sense, offering a unique window into the dynamics of cosmic baryons in this environment.

\subsection{Conclusions and future work}

Our work significantly advances our understanding of the magnetised Intragroup Medium (IGrM) and its extension into the cosmic web. Using data from the Polarisation Sky Survey of the Universe's Magnetism (POSSUM), we analyzed 22,817 Residual Faraday rotation measures (RRMs) across 55 diverse galaxy groups with halo masses between $10^{12.5}$ and $10^{14.0}$ M$_\odot$. Our key findings are:

\begin{itemize}
\item An Extended Magnetised IGrM: We confirm the presence of a magnetised IGrM extending up to twice the splashback radius for most galaxy groups, complemented by a secondary gas component—the Warm-Hot Intergalactic Medium (WHIM)—extending from 2 to 7 splashback radii. These components enhances RM dispersion by $6.9\pm1.8$ \radmm within the IGrM core and $4.2\pm 1.2$ \radmm in the WHIM, relative to previously established RRM dispersion contributions produced local to the RM sources of 6 \radmm. This discovery aligns with and expands upon recent theories and simulations \emph{vis-a-vis} the likely spatial distribution of `missing baryons', and suggests that a broader spatial distribution of ionised and magnetised gas around galaxy groups exists than has previously been recognised.

\item A High Prevalence of Magnetised IGrM: A large fraction of our sample groups contribute to the RRM dipserion inside the splahsback radius, and thus show evidence of a magnetised IGrM. This detection rate appears to be higher than X-rays currently suggest.

\item Diagnostic Value of RMs: Faraday Rotation Measures (RMs) have proven to be effective at detecting gas in low-density environments surrounding group-mass halos, pushing down in mass from the rich clusters that have more typically been used to probe. Beyond measuring magnetic fields, RMs can illuminate the structure and extent of the magnetised thermal IGrM and the cosmic web.

\item Implications for Cosmic Web Studies: The detection of magnetised gas extending well beyond the groups' splashback radii into the cosmic web underscores the dynamic state of galaxy groups, either accreting onto or being expelled from these structures. We appear to trace a denser phase of gas than previous studies focused on WHIM filaments \citep{Carretti2023}. These may represent the magnetised bubbles driven by galactic outflows that appear in cosmological simulations. 

\item Our results bridge the ranges of density and environments where RM grids have previously been found to provide useful constraints on magnetised gas properties. These experiments can now probe magnetised gas from galactic scale (e.g. \citealp{Heesen2023}) to groups (the present work), poor clusters \citep{Anderson2021} and rich clusters (e.g. \citealp{Clarke2001}), and beyond these structures into surrounding magnetised bubbles (present work, also \citealp{Bondarenko2024}) to sparse cosmic web filaments (e.g. \citealp{Carretti2023,BN2024}). RMs therefore appear to provide excellent tracers of magnetised gas throughout the cosmic cascade of baryons, from the largest scales down to galaxies.

\end{itemize}

Our study does have limitations: the RM data are not evenly distributed across galaxy groups and do not cover all scaled separations from each group. However, this analysis is based on data that covers only $\sim5$\% of the $2\pi$ steradians of sky that POSSUM will survey in the next few years with 30--50 RMs per square degree \citep{Vanderwoude2024}, Gaensler \emph{et al. in preparation}, and only $\sim3$\% of the $3\pi$ steradians that SPICE-RACS will observe with $\sim5$ RMs per square degree this year \citep{Thomson2023}. Consequently, the number of available RMs will greatly increase in the next few years, and sky coverage will expand dramatically over an even shorter timescale. While these improvements will not completely eliminate our current challenges, they will enhance the overall quality of RM grid experiments by increasing RM grid density and improving the uniformity and comprehensiveness of coverage. This will also enable more complex experiments in the future, such as examining magnetised gas in galaxy groups categorised by various properties.

Further advances will come from combining RM data with multi-wavelength observations such as X-ray emission, the Sunyaev-Zel'dovich Effect, and Fast Radio Burst dispersion measures. This will help clarify the role of magnetic fields in the cosmic baryon cycle around galaxy groups and how they influence the flow of energy and materials through this environment. The Square Kilometre Array (SKA), the Next Generation Very Large Array (ngVLA), and DSA-2000 will also in future provide denser RM grids through more sensitive measurements, broadening the scope of RM grid experiments as probe of group environments. In all of this, comprehensive magnetohydrodynamic  simulations that predict magnetic field properties and RM signals in group environments will be critical, providing a theoretical framework to interpret observational data, and which can be refined through RM data to constrain feedback models (e.g. \citealp{BN2024,Bondarenko2024}).

\section*{acknowledgement}

We thank the reviewer, Russ Taylor, for his time and comments, which have improved several important aspects of this work.

This scientific work uses data obtained from Inyarrimanha Ilgari Bundara / the Murchison Radio-astronomy Observatory. We acknowledge the Wajarri Yamaji People as the Traditional Owners and native title holders of the Observatory site. CSIRO’s ASKAP radio telescope is part of the Australia Telescope National Facility \footnote{\url{https://ror.org/05qajvd42}}. Operation of ASKAP is funded by the Australian Government with support from the National Collaborative Research Infrastructure Strategy. ASKAP uses the resources of the Pawsey Supercomputing Research Centre. Establishment of ASKAP, Inyarrimanha Ilgari Bundara, the CSIRO Murchison Radio-astronomy Observatory and the Pawsey Supercomputing Research Centre are initiatives of the Australian Government, with support from the Government of Western Australia and the Science and Industry Endowment Fund. The POSSUM project has been made possible through funding from the Australian Research Council, the Natural Sciences and Engineering Research Council of Canada, the Canada Research Chairs Program, and the Canada Foundation for Innovation. The Dunlap Institute is funded through an endowment established by the David Dunlap family and the University of Toronto. This work was partially funded by the Australian Government through an Australian Research Council Australian Laureate Fellowship (project number FL210100039) to N.Mc-G.. B.M.G. acknowledges the support of the Natural Sciences and Engineering Research Council of Canada (NSERC) through grant RGPIN-2015-05948, and of the Canada Research Chairs program. CIRADA is funded by a grant from the Canada Foundation for Innovation 2017 Innovation Fund (Project 35999), as well as by the Provinces of Ontario, British Columbia, Alberta, Manitoba and Quebec. TA was supported in part by JSPS KAKENHI Grant Number, JP21H01135. SPO acknowledges support from the Comunidad de Madrid Atracción de Talento program via grant 2022-T1/TIC-23797.

\section*{Data Availability}

This paper uses raw data products that are publicly available on the CSIRO ASKAP Science Data Archive (CASDA; \citealp{Chapman2017,Huynh2020}). A table of our RM data and galaxy group associations is available at CDS via anonymous ftp to cdsarc.u-strasbg.fr (130.79.128.5) or via https://cdsarc.cds.unistra.fr/viz-bin/cat/J/MNRAS at J/MNRAS/Vol/Page upon publication.



\bibliographystyle{mnras}
\bibliography{bibliography} 

\onecolumn
\begin{landscape}
\begin{longtable}{ccccccccccl}
\caption{Galaxy groups in present work} \label{tab:groups} \\
\hline
Tully Nest ID & NCG/IC/Messier & PGC & RA & Decl. & N Members & Splashback Radius & Splashback Angular Radius & Distance & Halo Mass & Parent SBID \\

 & & & [Deg.] & [Deg.] & & [Mpc] & [Deg.] & [Mpc] & $\mathrm{M_{\odot}}$ &  \\
\hline
\endfirsthead
\caption[]{(continued)} \\
\hline
Tully Nest ID & NCG/IC/Messier & PGC & RA & Decl. & N Members & Splashback Radius & Splashback Angular Radius & Distance & Halo Mass & Parent SBID \\

 & & & [Deg.] & [Deg.] & & [Mpc] & [Deg.] & [Mpc] & $\mathrm{M_{\odot}}$ &  \\
\hline
\endhead

200705 &  & 2443 & 10.193 & -38.343 & 4 & 0.68 & 1.11 & 70.0 & 2.5 & 46971 \\
201359 & NGC238 & 2595 & 10.857 & -50.183 & 3 & 0.79 & 1.04 & 87.0 & 4.0 & 46951 \\
200600 &  & 2948 & 12.674 & -52.22 & 4 & 0.75 & 1.07 & 80.0 & 3.4 & 46951 \\
201311 & NGC409 & 4132 & 17.388 & -35.806 & 3 & 0.6 & 1.06 & 65.0 & 1.7 & 46986 \\
200348 & NGC527 & 5128 & 20.992 & -35.115 & 6 & 0.69 & 1.39 & 56.0 & 2.6 & 46986 \\
200258 & NGC549 & 5278 & 21.369 & -38.267 & 7 & 0.75 & 1.44 & 60.0 & 3.4 & 46986 \\
201129 & NGC568 & 5468 & 21.988 & -35.718 & 3 & 0.58 & 1.22 & 55.0 & 1.6 & 46986 \\
200165 &  & 6104 & 24.744 & -46.573 & 9 & 0.85 & 1.51 & 65.0 & 5.0 & 46966 \\
201241 &  & 6131 & 24.853 & -49.382 & 3 & 0.71 & 0.89 & 91.0 & 2.9 & 46966, 47136 \\
201112 &  & 6994 & 28.302 & -49.56 & 3 & 0.72 & 1.32 & 62.0 & 3.0 & 47136 \\
200261 & NGC745 & 7054 & 28.532 & -56.694 & 7 & 0.78 & 1.51 & 59.0 & 3.8 & 46946 \\
200327 &  & 3628333 & 29.418 & -57.79 & 6 & 0.78 & 1.42 & 63.0 & 3.8 & 46946 \\
200448 &  & 8012 & 31.515 & -55.194 & 5 & 0.69 & 1.3 & 61.0 & 2.7 & 46946 \\
201362 & NGC822 & 8055 & 31.663 & -41.157 & 3 & 0.44 & 0.96 & 52.0 & 0.68 & 50415 \\
200506 & NGC835 & 8228 & 32.352 & -10.136 & 5 & 0.6 & 1.82 & 37.0 & 1.7 & 33370 \\
200771 & NGC842 & 8258 & 32.462 & -7.762 & 4 & 0.46 & 1.41 & 37.0 & 0.77 & 33370 \\
200984 &  & 8311 & 32.633 & -53.837 & 3 & 0.5 & 0.95 & 60.0 & 1.0 & 46946, 47034, 47136 \\
201282 &  & 8341 & 32.728 & -39.366 & 3 & 0.53 & 1.19 & 52.0 & 1.2 & 50415 \\
200783 &  & 8581 & 33.608 & -7.368 & 4 & 0.46 & 1.13 & 47.0 & 0.81 & 33370 \\
201051 & NGC883 & 8841 & 34.772 & -6.791 & 3 & 0.7 & 1.59 & 50.0 & 2.8 & 33370, 33460, 33482, 33509, 33553 \\
200640 &  & 9124 & 36.095 & -58.397 & 4 & 0.96 & 1.18 & 93.0 & 7.1 & 46946, 49990 \\
200271 & NGC945 & 9426 & 37.156 & -10.539 & 7 & 0.59 & 1.52 & 44.0 & 1.7 & 33482, 33553 \\
200602 &  & 9585 & 37.789 & -57.918 & 4 & 1.01 & 1.23 & 94.0 & 8.4 & 49990 \\
200740 &  & 9800 & 38.602 & -10.844 & 4 & 0.45 & 1.16 & 44.0 & 0.73 & 33482, 33553 \\
300025 & NGC 1068 &  & 38.866 & -9.356 & 22 & 0.6 & 3.5 & 20.0 & 1.8 & 33482, 33553 \\
201187 & IC243 & 10009 & 39.634 & -6.902 & 3 & 0.55 & 0.91 & 70.0 & 1.4 & 33460, 33482, 33509, 33553 \\
200685 & IC247 & 10100 & 40.037 & -11.734 & 4 & 0.75 & 1.05 & 82.0 & 3.4 & 33482, 33553 \\
200541 & NGC1045 & 10129 & 40.121 & -11.277 & 5 & 0.57 & 1.44 & 45.0 & 1.5 & 33482, 33553 \\
200503 & NGC1244 & 11659 & 46.63 & -66.776 & 5 & 0.66 & 1.34 & 56.0 & 2.3 & 46982 \\
200663 &  & 12231 & 49.384 & -54.358 & 4 & 0.87 & 1.16 & 86.0 & 5.3 & 50011 \\
200445 &  & 3095684 & 66.806 & -62.781 & 5 & 0.61 & 1.25 & 56.0 & 1.8 & 50230 \\
201088 &  & 15172 & 67.0 & -47.913 & 3 & 0.57 & 1.31 & 50.0 & 1.5 & 51431 \\
200155 &  & 16971 & 78.65 & -61.481 & 9 & 0.73 & 1.7 & 49.0 & 3.1 & 50538 \\
300277 & NGC 5078 & 0046304 & 199.428 & -32.102 & 4 & 0.38 & 1.51 & 29.0 & 0.44 & 43137 \\
300279 & NGC 5078 & 0046490 & 199.958 & -27.41 & 26 & 0.76 & 3.42 & 25.0 & 3.5 & 43137 \\
100394 & NGC5094 & 46580 & 200.195 & -14.081 & 5 & 0.86 & 1.41 & 70.0 & 5.2 & 31375 \\
100392 &  & 46747 & 200.737 & -32.728 & 5 & 0.97 & 1.23 & 90.0 & 7.4 & 43137 \\
100074 & NGC5135 & 46974 & 201.434 & -29.834 & 13 & 0.9 & 2.3 & 45.0 & 5.9 & 43137 \\
100391 &  & 47023 & 201.571 & -19.635 & 7 & 0.96 & 1.4 & 78.0 & 7.1 & 31375 \\
100341 &  & 48161 & 204.489 & -30.932 & 6 & 0.55 & 1.35 & 47.0 & 1.4 & 43137 \\
100179 &  & 50096 & 210.899 & -25.428 & 8 & 0.99 & 1.64 & 69.0 & 7.8 & 50413 \\
101226 &  & 715794 & 212.255 & -30.559 & 3 & 0.62 & 0.95 & 74.0 & 1.9 & 50413 \\
101276 &  & 184531 & 212.927 & -30.233 & 3 & 0.71 & 0.91 & 89.0 & 2.9 & 50413 \\
100753 &  & 50762 & 213.234 & -28.774 & 4 & 0.7 & 0.97 & 83.0 & 2.8 & 50413 \\
100361 & NGC5626 & 51794 & 217.454 & -29.748 & 6 & 0.88 & 1.41 & 71.0 & 5.5 & 46943 \\
101029 & IC4453 & 52084 & 218.619 & -27.519 & 3 & 0.52 & 1.5 & 40.0 & 1.2 & 46943 \\
100993 &  & 738224 & 219.365 & -28.724 & 3 & 0.62 & 1.04 & 68.0 & 2.0 & 46943 \\
201214 &  & 3902372 & 283.404 & -56.719 & 3 & 0.48 & 0.91 & 61.0 & 0.9 & 51574 \\
300356 & IC4797 & 0062589 & 284.124 & -54.306 & 8 & 0.65 & 2.04 & 36.0 & 2.2 & 51574 \\
200982 &  & 62713 & 285.823 & -53.206 & 3 & 0.72 & 0.91 & 91.0 & 3.0 & 51574 \\
200757 &  & 62721 & 285.914 & -53.934 & 4 & 0.55 & 1.36 & 47.0 & 1.4 & 51574 \\
200083 & NGC6753 & 62870 & 287.848 & -57.05 & 13 & 0.8 & 2.74 & 33.0 & 4.1 & 51574 \\
200765 & NGC6812 & 63625 & 296.351 & -55.347 & 4 & 0.56 & 1.42 & 45.0 & 1.4 & 51818 \\
201283 & IC4994 & 64489 & 304.935 & -53.447 & 3 & 0.54 & 1.09 & 57.0 & 1.3 & 51818 \\
201240 &  & 67429 & 327.453 & -59.323 & 3 & 0.76 & 1.05 & 83.0 & 3.6 & 45761 \\
\end{longtable}
\end{landscape}
\twocolumn

\bsp	
\label{lastpage}
\end{document}